\documentclass[sigconf]{acmart}

\AtBeginDocument{%
  \providecommand\BibTeX{{%
    \normalfont B\kern-0.5em{\scshape i\kern-0.25em b}\kern-0.8em\TeX}}}

\usepackage{algorithmic}
\usepackage{graphicx}
\usepackage{tcolorbox}
\usepackage{textcomp}
\usepackage{xcolor}
\usepackage{makecell}
\usepackage{enumitem}
\usepackage{multirow}
 
\usepackage{xcolor}
\usepackage{tikz}
\usepackage{pgfplots}
\usepackage{stfloats}
\usepackage{balance}

\definecolor{jspwiki}{HTML}{D7191C}
\definecolor{archiva}{HTML}{FDAE61}
\definecolor{jackrabbit}{HTML}{ABDDA4}
\definecolor{systemds}{HTML}{2B83BA}
\definecolor{helix}{HTML}{CB83BA}

\copyrightyear{2022}
\acmYear{2022}
\setcopyright{acmcopyright}\acmConference[ICPC '22]{30th International Conference on Program Comprehension}{May 16--17, 2022}{Virtual Event, USA}
\acmBooktitle{30th International Conference on Program Comprehension (ICPC '22), May 16--17, 2022, Virtual Event, USA}
\acmPrice{15.00}
\acmDOI{10.1145/3524610.3528387}
\acmISBN{978-1-4503-9298-3/22/05}

\begin{document}

\title{A First Look at Duplicate and Near-duplicate Self-admitted Technical Debt Comments}

\author{Jerin Yasmin}
\email{{jerin.yasmin, sadegh.sheikhaei, y.tian}@queensu.ca}
\author{Mohammad Sadegh Sheikhaei}

\author{Yuan Tian}

\affiliation{%
  \institution{Queen's University}
  \country{Canada}
}



\begin{abstract}

Self-admitted technical debt (SATD) refers to technical debt that is intentionally introduced by developers and explicitly documented in code comments or other software artifacts (e.g., issue reports) to annotate sub-optimal decisions made by developers in the software development process. 

In this work, we take the first look at the existence and characteristics of duplicate and near-duplicate SATD comments in five popular Apache OSS projects, i.e., JSPWiki, Helix, Jackrabbit, Archiva, and SystemML. We design a method to automatically identify groups of duplicate and near-duplicate SATD comments and track their evolution in the software system by mining the commit history of a software project. Leveraging the proposed method, we identified 3,520 duplicate and near-duplicate SATD comments from the target projects, which belong to 1,141 groups. We manually analyze the content and context of a sample of 1,505 SATD comments (by sampling 100 groups for each project) and identify if they annotate the same root cause. We also investigate whether duplicate SATD comments exist in code clones, whether they co-exist in the same file, and whether they are introduced and removed simultaneously. Our preliminary study reveals several surprising findings that would shed light on future studies aiming to improve the management of duplicate SATD comments. For instance, only 48.5\% duplicate SATD comment groups with the same root cause exist in regular code clones, and only 33.9\% of the duplicate SATD comment pairs are introduced in the same commit.

\end{abstract}



\maketitle

\section{Introduction}\label{sec:introduction}
Technical debt (TD) is a concept used to describe the trade-off between the ``quick and dirty'' design and implementation choices in modern software development and the long-term quality and maintainability of a software system~\cite{li2015}. Being aware of TD is crucial because it would introduce extra cost as TD accumulates over time due to sub-optimal implementation. Moreover, it would also increase the risk of performance issues and functionality bugs. 

\begin{figure}[htbp]
\captionsetup{skip=0pt}
    \centering
    \includegraphics[width=3.2in]{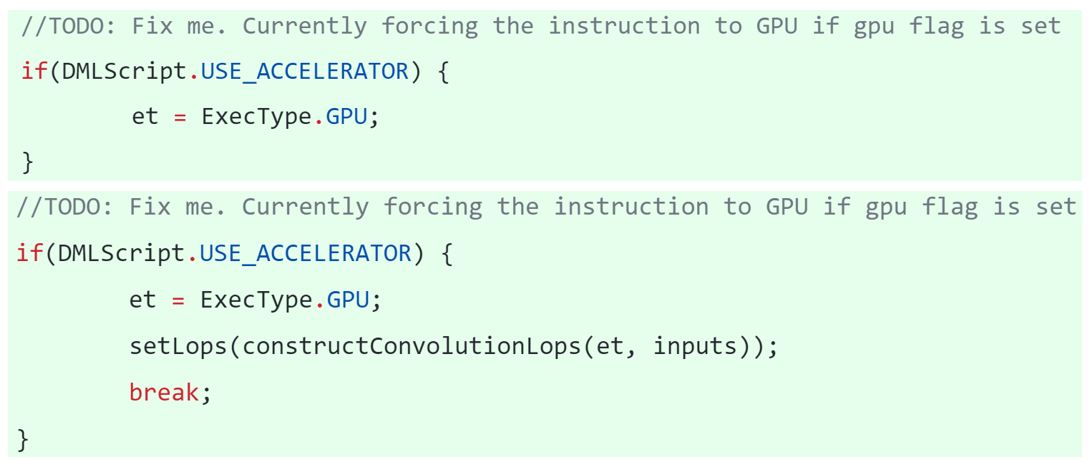}
    \caption{Example of duplicate SATD comments}
    \label{fig:sample}
\end{figure}

In recent years, researchers have intensively studied self-admitted technical debt (SATD), a concept proposed by Potdar and Shihab~\cite{potdar2014}, referring to the sub-optimal design and implementation decisions that developers explicitly acknowledge. SATD instances mainly appear as code comments~\footnote{SATD instances may also appear in issue report~\cite{xavier2020}.}, which we refer to as \textit{SATD comments}. Like general code comments, duplicate SATD comments may exist in a software system due to copy-pasting code snippets or annotating the same technical debt. Figure~\ref{fig:sample} shows an example pair of duplicate SATD comments in which the developer reports that the hard-coded assignment of a specific variable must be modified in the future to prevent bugs. However, most existing studies on SATD focus on identifying and tracking SATD instances. None of them have discussed the appearance of duplicate SATD comments. 



\vspace{-0.2cm}
\vspace{-0.3cm}

This paper conducts the first empirical study on the duplicate and near-duplicate SATD comments in open-source software projects. We define \textit{a group of duplicate and near-duplicate SATD comments} as a set of SATD comments with identical textual content or highly similar and semantically identical textual content. Each comment in the group is a duplicate or near-duplicate SATD comment, which we refer to as a \textit{\textbf{duplicate SATD comment}} in the rest of this paper. We aim to explore why duplicate SATD comments appear, where they appear and how they evolve in software systems. Answers to the above questions would help developers and researchers understand whether duplicate SATD comments would bring risks to the maintenance of projects and pose challenges in managing SATD instances. For instance, duplicate SATD comments might annotate the same bug and should be fixed together. If only one SATD instance is fixed, the remaining one might introduce a bug in the future. To achieve our goal, we develop a method to identify SATD comments from SmartSHARK~\cite{trautsch2021msr}, a dataset contains rich evolution information of 77 Apache projects. Among all the 77 projects, we picked five projects containing the most number of SATD comments in their commit history, i.e., JSPWiki~\footnote{https://jspwiki.apache.org/}, Helix~\footnote{https://helix.apache.org/}, Jackrabbit~\footnote{https://jackrabbit.apache.org/jcr/index.html}, Archiva~\footnote{https://archiva.apache.org/}, and SystemML~\footnote{https://systemds.apache.org/docs/1.2.0/}. Next, we applied a state-of-the-art natural language processing (NLP) model to compute semantic similarity scores between pairs of identified SATD comments from each target project and use them to identify groups of duplicate SATD comments. We use the collected data to answer the following three research questions:

\noindent \textbf{RQ1: Are duplicate SATD comments introduced by the same root cause? Do they exist in code clones?} Intuitively, duplicate SATD comments would be introduced for the same root cause, e.g., indicating areas in the code that are affected by the same (specific) poor implementation choice or the same bug. Also, many of duplicate SATD comments might appear when developers copy and paste code that contains a SATD comment. In RQ1, we test the above two hypotheses by manually examining a sample set of duplicate SATD comment groups. 


\noindent \textbf{RQ2: Are duplicate SATD comments exist in the same or different files?} For duplicate SATD comments in groups with the same root cause (identified in RQ1), we further check if they appear in the same file or different files. Intuitively, when developers resolve one SATD instance, they would be more likely to ignore the opportunities to change other code containing duplicate SATD comments if they appear in different files. Moreover, bugs could be introduced due to inconsistent changes on SATD instances annotated by duplicate SATD comments. 



\noindent \textbf{RQ3: Are duplicate SATD comments introduced and removed at the same time?} In RQ3, we investigate the evolution of duplicate SATD comment pairs with the same root cause (identified in RQ1). Answering this question would shed light on developers' current practice in introducing and evolving duplicate SATD comments.

\noindent \textit{Contribution.} We introduce a new concept, i.e., duplicate SATD comment. We gain the first preliminary understanding of duplicate SATD comments by analyzing thousands of SATD comments automatically mined from commit history of five Apache OSS projects. Our work would shed light on future analysis on the practice and quality of duplicate/general SATD documentation.

\noindent \textit{Data availability.} Our annotated duplicate and near-duplicate SATD comments are available at \url{https://doi.org/10.6084/m9.figshare.19426766.v1}.

\vspace{-0.1cm}
\section{Methodology}\label{sec:method}

\subsection{Data Collection and Target Projects}
We extract the SATDs from SmartSHARK (version 2.1) ~\cite{trautsch2021msr}, a dataset that contains detailed information about the evolution of 77 Apache projects that have Java as the main programming language. Given each project X in the dataset, we perform the following four steps to extract SATD comments by mining the whole commits in the master branch of project X:

\vspace{-0.1cm}
\begin{itemize}
    \item \textbf{Step 1:} Extract the sequence of commits that are in the master branch of the project X. 
    \item \textbf{Step 2:} Extract the ancestor of each file in project X by tracking the file renames in file\_action collection, and make the sequence of files for each final file (i.e. the files that have not been renamed anymore).
    \item \textbf{Step 3:} According to commits' order (step 1) and files' order (step 2), extract the ordered hunks from the hunk collection for each final file.
    \item \textbf{Step 4:} Scan the hunks of each final file in X to extract SATD. Following Trautsch et al.~\cite{trautsch2021msr}, we use task tags, i.e., ``TODO/FIXME/XXX'' to identify SATD from Java code comments. For each found SATD, we store its file\_id, file path, created\_in\_commit, created\_in\_line. Then, we track the changes in later hunks to find the final state of each found SATD. If a SATD comment gets deleted in later hunks, we also store the corresponding commit\_id.
\end{itemize}

After extracting SATD comments from the 77 projects in SmarkSHARK, we select the five projects that have the largest number of SATD comments (shown in Table~\ref{tab:data_collect}). Our target projects contain 7,861 SATD comments mined from the commit history of all five projects. Note that there might be SATD comments ignored by our approach, as they might use less explicit text to admit TD. However, as the first preliminary study on duplicate SATD comments, our goal is to understand their existence and characteristics, rather than reporting the precise ratio of duplicate SATD comments among all SATD comments. Moreover, Guo et al. found that detecting SATD comments by searching the appearance of predefined task tags ( ``TODO,'' ``FIXME,'' and ``XXX'') in code comments can achieve competitive performance when compared to complex machine learning-based approaches~\cite{guo2021far}.

\vspace{-0.2cm}
\subsection{Data Cleaning and Filtering}
Since task tags such as ``TODO'' do not carry meaningful information, we further clean extracted SATD comments by removing task tags. Then, we filter out the SATD comments that have the following properties:
\begin{itemize}
    \item Auto-generated SATD comments: They are not written by the developers (e.g., ``auto-generated method stub'', ``auto-generated catch block'').
    \item Single-word or numbers-only SATD comments: They are not informative to analyze the SATD they annotate.
\end{itemize}


\vspace{-0.2cm}
\subsection{Identifying Duplicate SATD comments} 

First, we leverage a pre-trained BERT (Bidirectional Encoder Representations from Transformers) model to generate a sentence embedding for each SATD comment. Specifically, we use the functions provided by SentenceTransformer\footnote{https://www.sbert.net/}. Next, we calculate the pairwise cosine similarity between each pair of SATD comment's embedding. We rank all pairs by their similarity scores and remove SATD comments that do not exist in any pair with a similarity score $\geq$ 0.8. The rational behind this threshold is that for the most pairs with a similarity score lower than 0.8, they are not duplicate SATD comments. For the remaining SATD comments, we generate a matrix $X$ where each row and each column represents a selected SATD comment, and the value of a cell $X_{ij}$ refers to the similarity score between SATD comment $i$ and $j$. We then fit the matrix $X$ as input to DBSCAN clustering algorithm~\cite{ester1996density} by calling the function \textit{fi\_predict} provided by package \textit{scikit-learn}~\footnote{https://scikit-learn.org/stable/}. We set the parameter ``eps" as 0.4 and ``min\_samples" as 2. 

In the end, we get a total of 1,141 groups containing 3,520 duplicate SATD comments (ref. Table~\ref{tab:data_collect}). \textbf{The ratio of the duplicate SATD comments that are found from the total number of SATDs are 65\%, 50.2\%, 50.3\%, 47.3\%, 41.3\% for project JSPWiki, Archiva, Jackrabbit, SystemML, and Helix respectively.}

\begin{table*}[t]
\caption{Basic dataset statistics. Filter 1 removes auto-generated SATD. Filter 2 removes single-word/number-only SATD.}
\vspace*{-3mm}
\resizebox{15.4cm}{!}{%
\centering
\begin{tabular}{|p{1.2cm}|p{3.2cm}|p{1.3cm}|p{2.3cm}|p{2cm}|p{2cm}|p{2cm}|p{2cm}|}
\hline
\textbf{Project } & \textbf{Domain} &\textbf{\# SATD} & \textbf{\# Removed by Filter 1} & \textbf{\# Removed by Filter 2}
& \textbf{\# SATD after filtering}& \textbf{\# Dup. SATD}& \textbf{\# Groups of Dup. SATD}\\
\hline
JSPWiki & Content, Wiki Engine & 1,325 & 243 & 15 & 1,067& 865&274\\
\hline
Archiva & Build Management & 1,508 & 327 & 61 & 1,120&757&241\\
\hline
Jackrabbit & Database & 1,721 & 407 & 119 & 1,195&725& 237\\
\hline
SystemML & Machine Learning & 1,559 & 281 & 51 & 1,227&737&232  \\
\hline
Helix & Big-data, Cloud & 1,748 & 1,056 & 15 & 677&436& 157\\
\hline
\end{tabular}
}
\label{tab:data_collect}
\end{table*}

\section{Preliminary Results}\label{sec:result}

\subsection{RQ1: Are duplicate SATD comments introduced by the same root cause? Do they exist in clones? }\label{sec:rq1}


\noindent \textbf{Approach:} To answer RQ1, we create a set that contains 500 duplicate SATD groups identified in Section~\ref{sec:method}. Specifically, we randomly sampled 100 groups from all identified groups for each target project. We choose to set the same number per project because we want to compare statistics across projects, and we do not want to bias a specific project. In total, these 500 groups involve 1,505 SATD comments. For each sampled group, for each duplicate SATD comment in the group, we carefully read the commit that introduced the SATD comment and identified the root cause of the TD this comment annotates. We label a group as \textit{a duplicate SATD comment group with identical root cause} (i.e., identical group) if all SATD comments in the group are introduced for the same reason and developers can potentially resolve them together.


To investigate the relationship between code clones and duplicate SATD comments, we first need to identify clones. We download the file associated with each target SATD comment and run NiCad~\cite{roy2008nicad} to identify block-level clones. Note that one SATD comment may exist in different versions of the same file, we consider the first version when the SATD was committed to code base. NiCad is a famous clone detection tool, as it has high precision and recall in identifying the code clones that are very similar but not an exact match, i.e., near-miss clones~\cite{roy2009mutation}. We set the similarity threshold of NiCad as 70\% (i.e., the default value) and the minimum length of a block at five lines, the best-reported value for clone detection using NiCad in Java source code~\cite{wang2013searching}. However, duplicate SATD comments might also exist in \textit{micro-clones}, which refers to the code clones having a size (LOC-line of code) smaller than the minimum size of regular code clones~\cite{mondai2018micro}. The size of a micro-clone ranges from 1 to 4 LOC. As micro-clones are difficult to detect due to the small size~\cite{mondal2020investigating}, we manually read the code surrounding (five lines of code above and below the line containing the SATD) of the SATD comment and determine if all SATDs in each group exist in the one micro-clones.

Two authors participate in the manual study. They first independently annotated 50 duplicate SATD comment groups. This process reported 4 groups with different and uncertain labels. The two authors discussed and resolved the conflict and one author then labeled the remaining 450 groups.

\vspace{.1cm}
\noindent \textbf{Results:} \textbf{484 out of 500 sampled duplicate SATD comment groups are groups with the identical root cause.} Among the 500 manually examined groups, we only find 16 groups containing duplicate SATD pairs that do not refer to the same problem, i.e., irrelevant. One potential reason is that most of our identified duplicate SATD comments are very specific in describing the corresponding TD. We manually checked the 16 non-identical groups and observed that most SATD comments in these groups are less informative, e.g., ``implement this", ``add more tests". That means SATD comments with general textual description would likely generate duplicate SATD comments referring to different specific problems and potentially introduce challenges to track and maintain SATD instances that should be or better be resolved together.

\textbf{48.5\% and 42\% of the duplicate SATD comment groups with the identical root cause belong to regular code clones and micro-clones.} The ratio in project JSPWiki, Archiva, Jackrabbit, SystemML, and Helix is 35\%, 53.6\%, 49\%, 60\%, and 45.3\% respectively. 
Our manual analysis finds that another 42\% (204 out of 484 groups) groups are within micro-clones. The ratio is 58\%, 35\%, 27.7\%, 38\%, 51\% for project JSPWiki, Archiva, Jackrabbit, SystemML, and Helix respectively. The results indicate the need in building micro-clone detection tool that can automatically detect and track the micro-clones containing duplicate SATD comments.

\begin{table*}[h]
\centering
\caption{Distribution of duplicate SATD comment pairs based on their introduction and removal status. Js/A/Ja/S/H refers to project JSPWiki/Archiva/Jackrabbit/SystemML/Helix.}
\label{tab:app-sk}
\vspace*{-3mm}
\resizebox{16cm}{!}{%
\begin{tabular}{|c|c|lllll|lllllllllllllll|}
\hline
\multicolumn{1}{|l|}{}                                                         & \multicolumn{1}{l|}{}                                                                      & \multicolumn{5}{l|}{}                                                                                   & \multicolumn{15}{c|}{Removed Status}                                                                                                                                                                                                                                                                                                                                  \\ \hline
\multicolumn{1}{|l|}{}                                                         & \multicolumn{1}{l|}{}                                                                      & \multicolumn{5}{c|}{\#pairs}                                                                            & \multicolumn{5}{c|}{\begin{tabular}[c]{@{}c@{}}Removed in\\ the same commits\\ or both remain\end{tabular}}                  & \multicolumn{5}{c|}{\begin{tabular}[c]{@{}c@{}}Removed in\\ different commits\end{tabular}}                                  & \multicolumn{5}{c|}{\begin{tabular}[c]{@{}c@{}}One is removed\\ and the other remains\end{tabular}}     \\ \hline
\multirow{6}{*}{\begin{tabular}[c]{@{}c@{}}Introduction\\ Status\end{tabular}} & \multirow{3}{*}{\begin{tabular}[c]{@{}c@{}}Introduced in\\ the same commit\end{tabular}}   & \multicolumn{5}{c|}{848}                                                                                & \multicolumn{5}{c|}{651}                                                                                                     & \multicolumn{5}{c|}{122}                                                                                                     & \multicolumn{5}{c|}{75}                                                                                 \\ \cline{3-22} 
                                                                               &                                                                                            & \multicolumn{1}{l|}{Js} & \multicolumn{1}{l|}{A} & \multicolumn{1}{l|}{Ja} & \multicolumn{1}{l|}{S} & H & \multicolumn{1}{l|}{Js} & \multicolumn{1}{l|}{A} & \multicolumn{1}{l|}{Ja} & \multicolumn{1}{l|}{S} & \multicolumn{1}{l|}{H} & \multicolumn{1}{l|}{Js} & \multicolumn{1}{l|}{A} & \multicolumn{1}{l|}{Ja} & \multicolumn{1}{l|}{S} & \multicolumn{1}{l|}{H} & \multicolumn{1}{l|}{Js} & \multicolumn{1}{l|}{A} & \multicolumn{1}{l|}{Ja} & \multicolumn{1}{l|}{S} & H \\ \cline{3-22} 
                                                                           &                                                                                            & \multicolumn{1}{l|}{61}   & \multicolumn{1}{l|}{170}  & \multicolumn{1}{l|}{251}   & \multicolumn{1}{l|}{271}  &  85 & \multicolumn{1}{l|}{17}   & \multicolumn{1}{l|}{136}  & \multicolumn{1}{l|}{201}   & \multicolumn{1}{l|}{229}  & \multicolumn{1}{l|}{68}  & \multicolumn{1}{l|}{4}   & \multicolumn{1}{l|}{28}  & \multicolumn{1}{l|}{40}   & \multicolumn{1}{l|}{46}  & \multicolumn{1}{l|}{4}  & \multicolumn{1}{l|}{40}   & \multicolumn{1}{l|}{6}  & \multicolumn{1}{l|}{10}   & \multicolumn{1}{l|}{6}  &  13 \\ \cline{2-22} 
                                                                               & \multirow{3}{*}{\begin{tabular}[c]{@{}c@{}}Introduced in\\ different commits\end{tabular}} & \multicolumn{5}{c|}{1,657}                                                                              & \multicolumn{5}{c|}{276}                                                                                                     & \multicolumn{5}{c|}{930}                                                                                                   & \multicolumn{5}{c|}{451}                                                                                \\ \cline{3-22} 
                                                                               &                                                                                            & \multicolumn{1}{l|}{Js} & \multicolumn{1}{l|}{A} & \multicolumn{1}{l|}{Ja} & \multicolumn{1}{l|}{S} & H & \multicolumn{1}{l|}{Js} & \multicolumn{1}{l|}{A} & \multicolumn{1}{l|}{Ja} & \multicolumn{1}{l|}{S} & \multicolumn{1}{l|}{H} & \multicolumn{1}{l|}{Js} & \multicolumn{1}{l|}{A} & \multicolumn{1}{l|}{Ja} & \multicolumn{1}{l|}{S} & \multicolumn{1}{l|}{H} & \multicolumn{1}{l|}{Js} & \multicolumn{1}{l|}{A} & \multicolumn{1}{l|}{Ja} & \multicolumn{1}{l|}{S} & H \\ \cline{3-22} 
                                                                               &                                                                                            & \multicolumn{1}{l|}{267}   & \multicolumn{1}{l|}{272}  & \multicolumn{1}{l|}{362}   & \multicolumn{1}{l|}{354}  & 402  & \multicolumn{1}{l|}{41}   & \multicolumn{1}{l|}{70}  & \multicolumn{1}{l|}{50}   & \multicolumn{1}{l|}{39}  & \multicolumn{1}{l|}{76}  & \multicolumn{1}{l|}{100}   & \multicolumn{1}{l|}{138}  & \multicolumn{1}{l|}{261}   & \multicolumn{1}{l|}{274}  & \multicolumn{1}{l|}{157}  & \multicolumn{1}{l|}{126}   & \multicolumn{1}{l|}{64}  & \multicolumn{1}{l|}{51}   & \multicolumn{1}{l|}{41}  & 169  \\ \hline
\end{tabular}
}
\label{tab:data_rq3}
\end{table*}
\subsection{RQ2: Are duplicate SATD comments exist in the same or different files? }\label{sec:rq2}

\noindent \textbf{Approach:} For each identical group of duplicate SATD comments $c_k$ that contains $n$ duplicate SATD comments, we first generate all unique duplicate SATD comment pairs $<s_i, s_j>$, where $s_i$ and $s_j$ are SATD comments in group $c_k$. A group with $n$ SATD comments would have $n(n-1)/2$ duplicate pairs. Next, for each $<s_i, s_j>$ pair, we identify if the two SATD comments exist in the same file or different files. If the group $c_k$ has $m$ duplicate pairs exist in the same file, we then calculate a ratio for group $c_k$, $\frac{m}{n(n-1)/2}$, denotes the prevalence of duplicate SATD comment pairs that exist in the same file in the group. 

\vspace{.1cm}
\noindent \textbf{Results:} \textbf{On average, 12.9\% to 72.1\% of the pairs in identical groups appear in the same file.} The 484 identical groups contain a total of 1,437 duplicate SATD comments, forming 2,505 duplicate SATD comment pairs. 
We observe that the ratio of duplicate SATD comment pairs in the same file varies among groups and across projects. For instance, the number of groups that have all SATD comments in the same file is 62, 25, 27, 57, 9, for project JSPWiki, Archiva, Jackrabbit, SystemML, and Helix respectively. We further perform four Mann-Whitney test~\cite{wilcoxon1992individual} comparing the group ratios between Helix and the other four projects. The test result shows that the groups from Helix project contain a significantly lower ratio of pairs in the same file ($p<0.05$). We then perform a manual examination on groups in Helix that have a low ratio of pairs appearing in the same file. One potential explanation is that many such pairs (in different files) are introduced because developers frequently change the project's file structure by removing one file and creating another file at a different location. This is understandable because Helix is a less mature project than the other four projects, with the latest version being v0.8.4 (in our dataset).


%


\subsection{RQ3: Are duplicate SATD comments introduced and removed at the same time?}\label{sec:rq3}

\noindent \textbf{Approach:} For each identical group of duplicate SATD comments (484 groups/1,437 SATD comments/2,505 pairs), we categorize the relationship of the two SATD comments in each pair of duplicate SATD comments by checking the commits that introduce the two comments and determine if both comments are introduced in the same commit. Next, we check if the two duplicate SATD comments have same removal status, i.e., removed in the same commit or both remain, remove in different commits, one is removed and the other remain in the system. The later two cases show inconsistent evolution of duplicate SATD comments. We perform the above analysis for pairs from each project and investigate if statistics vary across different projects. 


\vspace{.2cm}
\noindent \textbf{Results:} \textbf{848 (33.9\%) of the 2,505 studied duplicate SATD comment pairs are introduced at the same time. 77\% of the 848 pairs have a consistent removal status.} Table~\ref{tab:data_rq3} shows the distribution of different types of duplicate SATD comment pairs determined by their introduce and removal status. Surprisingly, across all 5 projects, less than 50\% of the considered pairs are introduced at the same time. This number varies when we break the numbers down to each project. For instance, the ratio of pairs introduced in the same commit ranges from 17.5\% to 43.4\%, with the highest ratio in project SystemML and the lowest ratio in project JSPWiki.
The ratio of pairs having consistent removal status are 27.8\%, 80\%, 80\%, 84.5\%, 80\%, for project JSPWiki, Archiva, Jackrabbit, SystemML, and Helix respectively.
For pairs that are introduced at the same commit but removed at different commit (122 pairs), we also measure the difference of their removal time and observe that the overall mean and median of the removal time difference is 8.3 days and 1.4 days respectively. This indicate potential time saving if we can improve the awareness of duplicate SATD comments. 




\vspace{-0.1cm}
\section{Related Work}\label{sec:relatedwork}
SATD was coined only a few years ago by Potdar and Shihab~\cite{potdar2014}, but many studies have been conducted, indicating the importance of this field of research. Sierra et al.~\cite{sierra2019} provided a survey of self-admitted technical debt and classified SATD research work into three categories: SATD detection~\cite{potdar2014,huang2018identifying,liu2018satd,guo2021far,maipradit2020}, SATD comprehension~\cite{li2015,xavier2020,bavota2016,zampetti2021,fucci2021waiting}, and SATD repayment~\cite{mensah2016rework,mensah2018}. In this regard, our study took place in SATD comprehension category. One close related work is from Gao et al.~\cite{gao2021automating}. They are concerned about the quality of SATD comments, focusing on obsolete SATD comments, and they proposed an approach that can automatically detect obsolete TODO comments.

\vspace{-0.1cm}
\section{Conclusion and Future Work}\label{sec:conclusion}
This paper introduces a new concept, i.e., duplicate SATD comment, referring to a SATD comment that is identical or highly similar (semantically identical) to one or more other SATD comments. We conduct the first empirical study by automatically identifying general and duplicate SATD comments from five Apache projects. We find that 41.3\%-65\% of our identified SATD comments are duplicate SATD comments. We manually analyzed 500 groups of duplicate SATD comments and found that the majority of them are identical groups, i.e., all SATD comments are introduced by the same root cause. We also find that 12.9\%-72.1\% of the pairs in identical groups appear in the same file, but only 33.9\% of the duplicate SATD comment pairs in identical groups are introduced in the same commit. In the future, we would like to expand the scope of our study and further investigate who introduced/removed/modified duplicate SATD comments, why they have different evolution in the project history, and what are the impacts of those duplicate SATD comments on the quality and evolution of software projects.

\section*{Acknowledgement}\label{sec.ack}
We acknowledge the support of the Natural Sciences and Engineering Research Council of Canada (NSERC), [funding reference number: RGPIN-2019-05071].
\clearpage
\balance
\bibliographystyle{ACM-Reference-Format}
\bibliography{main}

\end{document}